\documentstyle[sprocl]{article}

\input{psfig}

\def\Journal#1#2#3#4{{#1} {\bf #2}, #3 (#4)}

\def\NPB{{\em Nucl. Phys.} B}
\def\PLB{{\em Phys. Lett.}  B}
\def\PRL{\em Phys. Rev. Lett.}

\newcommand{\la}[1]{\label{#1}}
\newcommand{\be}{\begin{equation}}
\newcommand{\ee}{\end{equation}}
\newcommand{\ba}{\begin{eqnarray}}
\newcommand{\ea}{\end{eqnarray}}
\newcommand{\bi}{\begin{itemize}}
\newcommand{\ei}{\end{itemize}}
\newcommand{\fig}{Fig.~}
\newcommand{\nr}[1]{(\ref{#1})}

\newcommand{\fr}[2]{{\frac{#1}{#2}}}

\renewcommand{\vec}[1]{{\bf #1}}

\newcommand{\eq}{Eq.\,}

\newcommand{\h}{\hspace*{4mm}}

\newcommand{\su}[1]{\mbox{SU(#1)}}
\newcommand{\re}{\mbox{Re\,}}
\newcommand{\etal}{{\it et al}}
\newcommand{\fourth}{\mbox{$\fr14$}}

\def\lsi{\raise0.3ex\hbox{$<$\kern-0.75em\raise-1.1ex\hbox{$\sim$}}}
\def\gsi{\raise0.3ex\hbox{$>$\kern-0.75em\raise-1.1ex\hbox{$\sim$}}}

\def\u1{\mbox{U(1)}}
\def\su2u1{\mbox{SU(2)$\times$U(1)}}

\begin{document}

\vspace*{-2cm}\hfill NORDITA-1999/38HE\\

\title{MAGNETIC FIELD ON LATTICE \u1-HIGGS
AND \su2u1-HIGGS THEORIES\footnotemark}

\author{K. RUMMUKAINEN}

\address{%
NORDITA, Blegdamsvej 17, DK-2100 Copenhagen \O, Denmark;\\ 
Helsinki Institute of Physics, P.O.Box 9, 00014 University of Helsinki, Finland\\
E-mail: rummukai@nbi.dk}


\maketitle\abstracts{%
External (hyper)magnetic field can modify the phase structure in \u1
gauge+Higgs (Landau-Ginzburg) and \su2u1 gauge+Higgs (Standard Model)
theories.  In this talk I discuss how the magnetic field can be
implemented on the lattice, and summarize the effects on symmetry
breaking phase transitions.}

\section{Introduction}

\footnotetext{Presented at the
ECT$^*$ conference {\em Understanding Deonfinement in QCD\,}, Trento, Italy, March 1999.}

Theories with a \u1 gauge field allow gauge invariant (global)
magnetic fields.  In this talk I discuss the Higgs field driven phase
transitions in 3-dimensional \u1 gauge+Higgs\cite{vortex} and 
\su2u1 gauge+Higgs\cite{ewpt}
theories, and especially what happens to the transitions when an
external magnetic field is applied\footnote{The research summarized
here has resulted from collaborations with K. Kajantie, M. Laine,
T. Neuhaus, J. Peisa and A. Rajantie (\u1 gauge+Higgs), and with K.K,
M.L, J.P, P. Pennanen, M. Shaposhnikov and M. Tsypin (\su2u1
gauge+Higgs).}.  As is appropriate for the topic of this workshop,
I shall point out some similar features to lattice QCD simulations
with a fixed baryon number.

Several reasons make \u1 gauge+Higgs theory with external magnetic
field interesting: first of all, it is the dimensionally reduced
version of 3+1-dim. scalar QED at high temperatures.\cite{vortex}  It
also is equivalent to the Ginzburg-Landau theory of superconductivity,
and it is very useful as a ``toy model'' for studying the behaviour of
string-like defects in cosmology (Nielsen-Olesen strings).  On the
other hand, 3d \su2u1 gauge+Higgs theory is precisely the
high-temperature effective theory of the Standard Model (and of many
of its extensions).  The physical consequences of the electroweak
phase transition in the early Universe can be substantially modified
in the presence of a hypermagnetic field (homogeneous in microscopic
scales).

\section{\u1+Higgs Theory with Magnetic Field}

The action of the \u1+Higgs theory in the continuum is
\be
 S = \int d^3 x \big[ \fourth F_{ij}F_{ij} + (D_i \phi)^*D_i\phi 
     + m_3^2 |\phi|^2 + \lambda_3 |\phi|^4 \big]\,,
\ee
where the couplings $e_3^2$, $m^2_3$ and $\lambda_3$ are 
dimensionful.  The lattice action (with non-compact \u1)
can be written as
\be
 S_L = \sum_x \bigg[ \beta_G \sum_{ij} \fr12 \alpha_{ij}^2 
	-\beta_H\sum_i \re(\phi^*_x U_{x,i} \phi_{x+i}) + 
	 \beta_2 |\phi|^2 + \beta_4 |\phi|^4 \bigg] \,,
\la{u1action}
\ee
where the plaquette $\alpha_{x,ij} = \alpha_{x,i} + \alpha_{x+i,j} - 
\alpha_{x+j,i} - \alpha_{x,j}$ and $U_{x,i} = \exp i \alpha_{x,i}$.
The gauge coupling is $\beta_G = 1/(e_3^2 a)$, and the relations of
$\beta_H$, $\beta_2$ and $\beta_4$ to the continuum parameters are
given in\cite{vortex}.  Note that the 3d theory is 
superrenormalizable, and it has a well-defined continuum limit.

How should one proceed in order to introduce an external magnetic
field in the system?  For definiteness, let us fix $\vec
H\parallel\vec B\parallel\hat e_3$.  The customary way of introducing
the field is to add a source term to the action:
\be
Z(H) = \int [dU d\phi] \exp \big[\mbox{$-S + \int dx  B(x) H$}\big] \,.
\la{canonical}
\ee
Here $B(x) = F_{12}(x) = \beta_G^2{e_3^3}\alpha_{12}(x)$ is the (local)
magnetic field.  However, this approach does not work on a
finite lattice with periodic boundary conditions: if we consider any
plane with a fixed $x_3$-coordinate, then $\sum_{x_1,x_2} \alpha_{12}(x) = 
\sum_{\partial A} \vec\alpha \cdot \hat e_{\partial A} = 0$, where
$\partial A$ is the boundary of the (1,2)-plane.  Thus, external field
$H$ has no effect on $Z$!\footnote{If one uses a {\em compact} gauge
action, the total flux can fluctuate by units of $2\pi$.  In practice,
the spontaneous fluctuations are extremely strongly suppressed, but it
is possible to construct a global update which substantially enhances
the fluctuations.\cite{Damgaard} It is straightforward to generalize
this update to a non-compact gauge action, but for our lattices the
fluctuations still remain too strongly suppressed.}

One method to solve this problem is to use modified boundary
conditions.\cite{vortex,ewpt}  For example, let us choose link
$\alpha_1(1,1,x_3)$ on each (1,2)-plane, and use a non-periodic
boundary condition for this link alone:
\be
  \alpha_1(1,1,x_3) = \alpha_1(1,N,x_3) + \Delta\,,
 \mbox{~~~otherwise $\alpha_i$ periodic.}
\la{nonperiod}
\ee
Here $L^3 = (aN)^3$ is the size of the lattice.  If, for the time being,
we neglect the Higgs field, this extra `twist' can
distribute itself evenly on all (1,2)-plaquettes: $\langle
\alpha_{12}\rangle =\Delta/N^2$, 
corresponding to a homogeneous magnetic field $B_3 =
\Delta/(e_3 L^2)$.
It should be noted that the particular choice of the link in 
\eq\nr{nonperiod} does {\em not} give that link a special status:
the action --- and physics --- remains translationally invariant.
If we now make $\Delta$ a dynamical variable,
the source term in \eq\nr{canonical} simply becomes $\Delta H/e_3$,
and canonical simulations become possible.

However, when the Higgs field is taken into account, a new condition
arises: the hopping term in \eq\nr{u1action} is translation invariant
only if the condition $\Delta = 2\pi n$, $n$ integer, is satisfied.
If this is not the case, there will be a localized `defect' on the
link $\alpha_1(1,1,z)$, and these configurations are unphysical.  In
principle, one may still attempt to update $n$ dynamically in integer
units, but these updates are extremely strongly suppressed.  In practice
the total flux $\Phi_B$ remains fixed, and we have an
ensemble
\be
  Z(\Phi_B) = \int [dU d\phi] e^{-S} \delta(\Phi_B(U) - \Phi_B)\,,
  \h \Phi_B = 2\pi n/e_3\,.
\la{microcanonical}
\ee 
In what follows I call this a microcanonical ensemble.  When one scans
the phase space of the system, this ensemble causes several problems.
For example, let us now consider the case $\lambda_3/e_3^2 \ll 1$,
which (in terms of Ginzburg-Landau theory) describes a type I
superconductor.  When $m_3^2$ (which is now a function of temperature
$T$) is decreased from positive values to negative ones, the system
has a strong 1st order phase transition from the symmetric to the
broken (Higgs) phase. In the canonical constant $H$ ensemble 
\eq\nr{canonical}, the magnetic field cannot penetrate the broken
phase, forcing $B=0$ due to Meissner effect, as shown schematically in
\fig\ref{fig:phdiag}.  However, in the microcanonical ensemble \nr{microcanonical}
the flux is fixed and cannot vanish.  The system can accommodate this
only by forming a {\em mixed phase}: the magnetic field penetrates the
volume through a cylinder with a small cross-sectional area, so that
the field strength inside the cylinder is large enough to support
unbroken phase at this temperature.

\begin{figure}[t]
\centerline{\psfig{figure=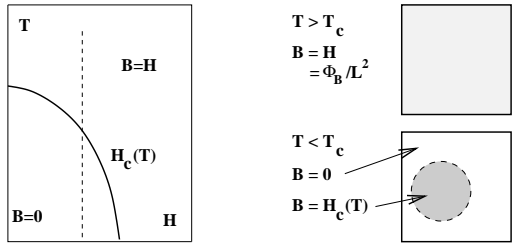,height=3.2cm}}
\caption[a]{{\em Left:} the phase diagram of type I superconductor
in $H,T\sim m_3^2$ plane.  {\em Right:} (1,2)-plane cross sections
of the system in the microcanonical ensemble \eq\nr{microcanonical}.
System goes in a mixed phase when $T < T_c$, where $H_c(T_c) = \Phi_B L^2$.
In the mixed phase the flux is concentrated in a subvolume with a
cross-sectional area $A$ so that $AB = AH_c(T) = \Phi_B$.}
\la{fig:phdiag}
\end{figure}

An immediate consequence of the existence of the mixed phase is the
`rounding off' of the phase transition: if we are in the mixed phase
and increase $T$, the symmetric phase domain grows smoothly, until at
$T=T_c$ it takes over the whole volume.  There are no
discontinuities in thermodynamical densities when the external
parameters ($\Delta$, couplings) are varied.  This is in strong
contrast to the canonical ensemble, which typically displays strong
discontinuities in 1st order phase
transitions.\footnote{Interestingly, the first order nature in the
microcanonical ensemble is recovered when $\phi_B=0$, which
corresponds to the standard simulation without any external fields.}
This property is a generic feature of almost any `microcanonical-like'
ensembles.  For example, a very analogous situation occurs in QCD
simulations with a fixed baryon number,\cite{Kaczmarek} where the
transition from hadronic phase to quark-gluon plasma becomes
continuous.

One way to work around the quantization condition in
\eq\nr{microcanonical} is to use {\em multicanonical\,} (or related)
methods to interpolate $\Delta = 2\pi n$ from $n$ to
$n+1$, for example.  Even when the non-integer values of $n$ do not correspond
to physical configurations, the free energy difference obtained from
the integral
\be
    \delta F(n\rightarrow n+1) = \int_{2\pi n}^{2\pi(n+1)} d\Delta\,  
	\fr{\partial F}{\partial \Delta}
\ee
is a fully physical quantity.  In type II domain $\lambda_3/e_3^2 >
1$, where the magnetic field in the Higgs phase forms unit flux
vortices, the integration method was succesfully used to measure the vortex
tension = free energy/length.\cite{vortex}  This, in turn, can be
used as an order parameter between the symmetric/broken phases.

\section{\su2u1-Higgs Theory}

3d \su2u1 gauge+Higgs theory is an effective theory of high $T$ Standard
Model.\cite{ewpt,fkrs} The Lagrangian of the theory is
\be
  \fourth F^a_{ij} F^a_{ij}  +
  \fourth B_{ij} B_{ij}  +
  (D_i \phi)^\dagger D_i\phi + m_3^2\phi^\dagger\phi + 
  \lambda_3(\phi^\dagger\phi)^2\,,
\la{lagr}
\ee
where $F_{ij}^a$ and $B_{ij}$ are SU(2) and U(1) field strength
tensors with gauge fields $A_i^a$ and $B_i$, the covariant derivative
$D_i = \partial_i + ig_3 A_i + ig'_3 B_i/2$, and the Higgs field
$\phi$ is a complex doublet.  Here the 3d gauge couplings $g_3^2
\approx g^2 T$, $g'_3{}^2 \approx g'{}^2 T$, and we fix $z =
g'{}^2_3/g^2_3 = 0.3$.  The results are conveniently expressed in
terms of dimensionless variables $x = \lambda_3/g_3^2$ and $y=
m_3^2/g_3^4$.

Since \eq\nr{lagr} contains the hypercharge \u1 field, we can study
the phase transition in the presence of an external hypermagnetic
field with similar methods as for the \u1-Higgs theory above.  As
before, the flux is quantized, and we are restricted to the
microcanonical ensemble as in \eq\nr{microcanonical}.  However, in the
present case the hypermagnetic field can penetrate the broken (Higgs)
phase as an ordinary magnetic field, albeit with a cost in free
energy.  Again, at small $x$ the transition is of 1st order, and in
the microcanonical ensemble we expect the single transition line to be
split into a band of mixed phase.  Using tree-level analysis, this is
shown in \fig\ref{fig:treel}.  At large $x$ there is a region where
neither symmetric nor broken Higgs phases are stable (again, at tree
level!), and it is possible that one finds here a new vortex-like
Ambj{\o}rn-Olesen\cite{Ambjorn} phase, which 
breaks translational invariance.

\begin{figure}[t]
\centerline{\psfig{figure=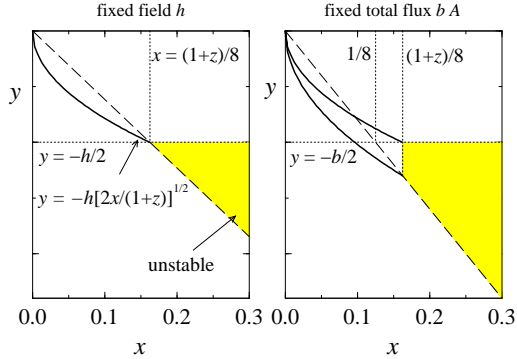,height=5cm}}
\vspace{-4mm}
\caption[a]{The tree-level phase diagram in \su2u1 gauge+Higgs theory
with fixed external field $H$ (canonical ensemble, left) and fixed
total flux $B\times$ area (microcanonical ensemble, right).  The
single first order transition line at small $x$ splits into a band
where a symmetric/broken mixed phase can be found.  The shaded region
at large $x$ shows where the Ambj{\o}rn-Olesen phase can be expected
to appear.\cite{Ambjorn}}
\la{fig:treel}
\end{figure}
\begin{figure}[t]
\centerline{\psfig{figure=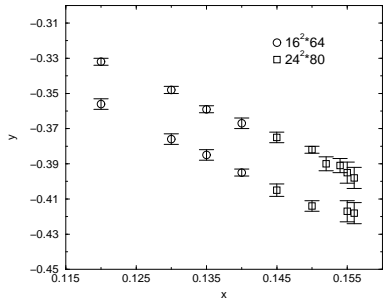,height=4cm}}
\vspace{-3mm}
\caption[a]{The measured mixed phase band when the flux
density is $B_Y \approx 8 g^4T^2/3g'$.} 
\la{fig:lattice}
\end{figure}

Indeed, in extensive lattice simulations we do observe a mixed phase,
as shown in Figs.\,\ref{fig:lattice} and \ref{fig:rope}.  Not
surprisingly, though, the quantitative differences from tree-level
results are large.  On the other hand, at larger $x$ we do not observe
the Ambj{\o}rn-Olesen phase.  Instead the broken and symmetric phases
appear to be smoothly connected in this region, compatible with the
observed behaviour in the absence of magnetic field.  For all of the
details, see refs.\cite{ewpt,prep}.

\begin{figure}[t]
\centerline{\psfig{figure=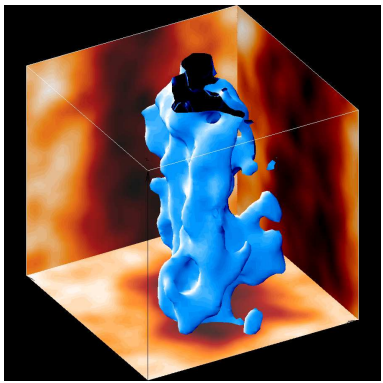,height=5cm}}
\caption[a]{A mixed phase configuration in \su2u1+Higgs theory.
The hypermagnetic flux preferentially goes through a cylinder of the
symmetric phase embedded in the bulk broken phase.}
\la{fig:rope}
\end{figure}

\section*{Acknowledgments}
I thank my collaborators for numerous discussions, and the organizers
for a very interesting conference.  The research was partially
supported by the TMR network {\em Finite Temperature Phase Transitions
in Particle Physics}, EU contract no.\ FMRX-CT97-0122.

\end{document}